\documentclass[prl,preprint,showpacs,amsmath,amssymb]{revtex4}
\usepackage{graphicx,amssymb}

\begin{document}
\title[Atomistic Simulations of the Mechanical Properties of ``Super'' Carbon Nanotubes]{Atomistic Simulations of the Mechanical Properties of ``Super'' Carbon Nanotubes}
\author{V R Coluci$\dag$, N M Pugno$\ddagger$, S O Dantas$\star$, D S Galv\~ao$\dag$, A Jorio$\star\star$}

\address{$\dag$ Instituto de F\'{\i}sica ``Gleb Wataghin'',
Universidade Estadual de Campinas, Unicamp 13083-970, Campinas,
S\~ao Paulo, Brazil}

\address{$\ddagger$ Department of Structural Engineering, Politecnico di Torino, Corso Duca degli Abruzzi 24, 10129 Torino, Italy}

\address{$\star$ Departamento de F\'{\i}sica, ICE, Universidade Federal de Juiz de Fora, 36036-330 Juiz de Fora MG, Brazil}

\address{$\star\star$ Departamento de F\'{\i}sica, Universidade Federal de Minas Gerais, 30123-970 Belo Horizonte MG,  Brazil}

%\ead{coluci@ifi.unicamp.br}
\date{\today}

\begin{abstract}
The mechanical properties of the so-called `super' carbon nanotubes (STs) are investigated using classical molecular dynamics simulations. The STs are built from single walled carbon nanotubes (SWCNTs) connected by Y-like junctions forming an ordered carbon nanotube network that is then rolled into a seamless cylinder. We observed that the ST behavior under tensile tests is similar to the one presented by fishing nets. This interesting behavior provides a way to vary the accessible channels to the inner parts of STs by applying external mechanical load. The Young's modulus is dependent on the ST chirality and it inversely varies with the ST radius. 
Smaller reduction of breaking strain values due to temperature increase is predicted for zigzag STs compared to SWCNTs. The results show that, for STs with radius $\sim$5 nm, the junctions between the constituent SWCNTs play an important role on the fracture process. The Young's modulus and tensile strength were estimated for hierarchical higher-order STs using scaling laws related to the ST fractal dimension. The obtained mechanical properties suggest that STs may be used in the development of new porous, flexible, and high-strength materials.

\end{abstract}

%\pacs{73.22.-f, 81.05.Tp, 81.05.Zx}

%\submitto{\NT}
\maketitle

\section{Introduction}

Many attempts have been made in order to develop procedures to controllably assemble large number of single walled carbon nanotubes (SWCNTs) in terms of position and orientation \cite{diehl,avouris,ago,ismach,net,ismach2}. The achievement of such procedures would allow the fabrication of ordered SWCNT networks representing a breakthrough in the ``bottom-up'' manufacturing approach. These ordered networks would open possibilities to design new materials with desirable electronic and mechanical properties. 

Recently, the structure of the so called `super' carbon nanotubes (STs) was proposed \cite{coluci} (Figure 1). This structure can be generated from an ordered carbon network based on the honeycomb symmetry, generically named super-graphene, which is heuristically constructed replacing the carbon-carbon bonds of the graphene architecture by single walled carbon nanotubes (SWCNTs) and the carbon atoms by Y-like junctions. The associated STs can be then generated by rolling up super-graphene sheets. Similarly to a ($n$,$m$) SWCNT \cite{dresstubes}, [$N$,$M$] ST with different chiralities can be constructed. The STs are represented as $[N,M]@(n,m)$ and characterized by the ($n$,$m$) SWCNT used to form them, the necessary junctions to join consecutive SWCNTs, and the distance between these junctions. The ST construction is not limited to carbon nanotubes and to the honeycomb symmetry, and it represents a three-dimensional network of nanotubes. It can be applied to other tubular structures and to different symmetries through the use of other junction types (X- or T-junctions, for example). Romo-Herrera \textit{et al.} has expanded this method of super structure construction to build two- and three-dimensional structures connecting SWCNTs with appropriate junctions to form super-square, super-cubic, and super-diamond structures \cite{terr}. 

Hierarchical structures \cite{lakes} can be build generalizing the procedure to construct STs leading to higher order hierarchical STs. The lowest order ST (ST$^{(0)}$) is the SWCNT, representing the building block for the next order, i.e., ST$^{(1)}$. A ST of order $k$ can be recursively made of STs of the previous order $k -1$ (ST$^{(k-1)}$ $\rightarrow$ ST$^{(k)}$) leading to a self-similar structure \cite{coluci}. The electronic structure of some ST$^{(1)}$s has been determined by tight-binding calculations \cite{coluci} showing a very rich (either metallic or semiconducting) behavior depending on the ($n$,$m$) SWCNT used in the ST construction. In this work we have used fully atomistic simulations to determine the mechanical properties of ST$^{(1)}$s under impact load. The derived mechanical properties, such as Young's modulus and tensile strength, were estimated for larger (e.g., higher-order) STs using scaling laws related to the ST fractal dimension \cite{pugno}. Using these laws it was predicted that the material efficiency (index of the optimization between strength and toughness) presents a maximum when the zigzag ST order is $k \simeq$ 2 \cite{pugno}, which is similar to the case of nacre \cite{nacre1,nacre2}. Recently Pugno \cite{pugno} has suggested the use of STs as hierarchical fibre-reinforcements embedded in a soft matrix for producing bioinspired synthetic `super'-composites.

A direct route to produce ST is beyond our present technological capabilities. Progress in synthesis of pure SWCNT with well defined diameter and chirality is also needed to allow the production of pure STs. Baughman \textit{et al.} have proposed a strategy based on topochemical reactions which can lead to the synthesis of specific SWCNTs \cite{topo}. Recent advances in the controlled fabrication of ordered carbon nanotubes networks \cite{diehl, avouris,ago,ismach,net, ismach2} can help on the development of techniques suitable to produce STs. It is encouraging that complex structures involving even more complex molecules, like DNA, can currently be produced \cite{dna}.

Mechanical properties of super-graphene have been studied using the finite element method \cite{wang}. The structures were composed by (15,15) SWCNTs of length of 20 nm arranged in a honeycomb configuration. These studies revealed that those structures have great flexibility and a remarkable capability of force transferring \cite{wang}. Fully atomistic simulations are not feasible for such huge structures modeled by finite element method, but they can be used to investigate networks formed  by shorter SWCNTs and with a higher density of junctions. In those cases, the changes of angles between SWCNTs during mechanical deformations can be larger than the ones observed in larger structures. We show here that these large angle changes are of fundamental importance in the rupture process during tensile deformations in STs.

%Figure 1
\begin{figure}
\begin{center}
\includegraphics[angle=0,width=120 mm]{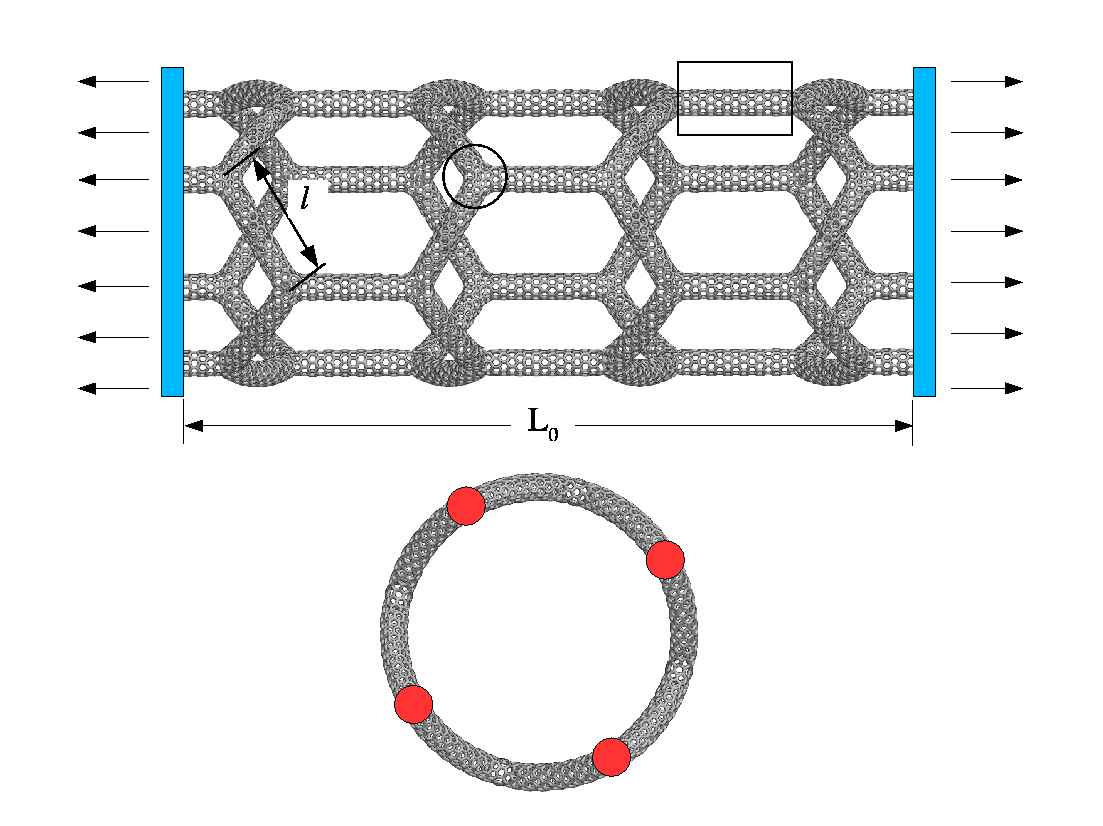}
\caption{Longitudinal (top) and cross section (bottom) views of a [4,0]@(8,0) `super' carbon nanotube. In the super carbon nanotube construction single walled carbon nanotubes (rectangular box) are connected by Y-like junctions (circle). The atoms in the super carbon nanotube extremities (blue boxes) are moved with a constant speed during tensile tests. Red disks represent the cross-sectional ST area.}
\end{center}
\end{figure}

\section{Methodology}

We have investigated zigzag [$N$,0]@(8,0) and armchair [$M$,$M$]@(8,0) STs with $4 \leq N \leq 6$ and $4 \leq M \leq 6$, with initial tube lengths $L_0$ ranging from 8.6 to 35.2 nm, corresponding to two ST unit cells. The radius $R_{[N,M]}$ of a [$N$,$M$]@($n$,$m$) is given by $R_{[N,M]}=l\sqrt{3(N^2+M^2+ NM)}/(2\pi)$, where $l$ is the distance between SWCNT junctions (Figure 1). The values of ST radii were in the range of 2.7 up to 16.9 nm. The (8,0) SWCNTs were connected using junctions constructed with 5- and 8-membered rings \cite{junc1}. Chiral STs ([4,1]@(8,0)) of length 11.4 and 15.3 nm, and radii 3.13 and 4.20 nm, respectively, were used for comparison purposes.

Classical molecular dynamics simulations were used to carry out ST tensile tests. The interactions between carbon atoms were described by the adaptive intermolecular reactive empirical bond-order potential developed by Stuart \textit{et al.} \cite{airebo}. This potential is similar to the reactive potential developed by Brenner \cite{brenner} but it incorporates by suitable modifications the non-bonded interactions through an adaptative treatment of the intermolecular interactions. In order to prevent the overestimation needed to break a carbon-carbon covalent bond, that it is a known problem occurring with Brenner potential due to the functional form of the cutoff function \cite{fracture, fracture1}, we have increased the covalent interaction cutoff distance from 1.7 to 1.95 {\AA} \cite{susan}. During the molecular dynamics simulations the Newton's equations of motion were integrated with a third-order Nordisieck predictor corrector algorithm \cite{nordsieck} using a time step of 0.5 fs. 

We have carried out simulations of tensile tests in situations of impact load, i.e., the atoms on the ST extremities were moved along the axial directions with a speed of 10 m/s (Figure 1). In order to investigate thermal effects on the tensile behavior of STs, the Berendsen's thermostat \cite{berendsen} was applied to all remaining atoms. Tensile behaviors for two different temperatures (300 K and 1000 K) were analyzed. From the initial stages of the tensile tests, i.e., for tensile strains $\epsilon=(L-L_0)/L_0$ ($L$ is the deformed ST length) less than 0.05, we have determined the axial dynamic Young's modulus $E$ using $E=(\partial^2 U/\partial\epsilon^2)n_p/(A_{[N,M]}L_0)$, where $n_p$ is the number of atoms, $A_{[N,M]}$ is the cross-sectional ST area where the tensile force is applied, and $U$ is the potential energy per atom of the system. For the zigzag STs considered in this work $A_{[N,0]}=N A_0$, where $A_0=2\pi r h$ is the cross-sectional (8,0) SWCNT area ($r=$ 3.13 {\AA}, $h=$ 3.4 {\AA} is the usual SWCNT shell thickness \cite{34}), and for armchair STs $A_{[M,M]}=2hlM$. Since the armchair ST radius is $R_{[M,M]}=3lM/(2\pi)$ then $A_{[M,M]}$ can also be written as $A_{[M,M]}=4\pi h R_{[M,M]}/3$. We have only estimated the cross-sectional area of chiral STs using data from the Connolly surfaces \cite{surface} since they did not present a simple closed form as in zigzag and armchair STs. We used the maximum value of the tensile stress as an estimate of the dynamic tensile strength $\sigma_m$ of the STs, and we defined the corresponding tensile strain as the breaking strain $\epsilon_m$. The breaking strain does not necessarily characterize the full breaking of STs but indicates their first local rupture. 

\section{Results}

Figure 2 presents the results of the Young's modulus obtained from atomistic calculations. We can see that the Young's modulus of STs decreases with the increase of the ST radius for fixed $N$ and $M$ values. This behavior can be expressed through the relation $E=E_0(R/R_0)^{-\beta}$ ($R\equiv R_{[N,M]}$) which reasonably fits the atomistic results as shown in Figure 2. We have chosen $R_0$ as being the radius of a given ST. In Table 1 are presented the fitted values of the parameters $E_0$ and $\beta$. The $E$ maximum value obtained was about 400 GPa for a [6,6]@(8,0) ST. For a fixed value of $R$, $E$ increases with the increasing of the $N$ and $M$ values. These results indicate that the ST stiffness can be increased by increasing the number of junctions (varying $N$ and $M$ for a fixed radius). On the other hand, ST flexibility can be enhanced by increasing the length $l$ of the constituent SWCNTs. For STs formed with longer SWCNTs, the high bending flexibility of these SWCNTs is transfered to the behavior of STs. The mechanical behavior is similar to one presented by a fishing-net. It is relatively easy to deform the net up to a large elongation. This happens due to the flexibility of net arms and the small bending modulus at the junctions. The elongation occurs until the junction angles are in their maximum apertures. After this point the mechanical behavior is due to the mechanical properties of the (stiff) arms themselves.  An investigation using finite element methods for ordered carbon nanotube networks formed by long SWCNTs (length of 20 nm) have revealed that such structures have indeed great flexibility \cite{wang}.

Using our procedure for tensile tests the Young's modulus for the (8,0) SWCNT was about 850 GPa. This value is in agreement with values obtained in other theoretical simulations using Brenner potential \cite{mielke} but it is about 15\% smaller than the reference value of 1 TPa obtained for SWCNTs \cite{young}. Thus, the absolute values derived for the mechanical properties of STs must be considered with caution and analyzed comparatively in terms of the values obtained for the constituent (8,0) SWCNT.

\begin{figure}
\begin{center}
\includegraphics[angle=0,width=120 mm]{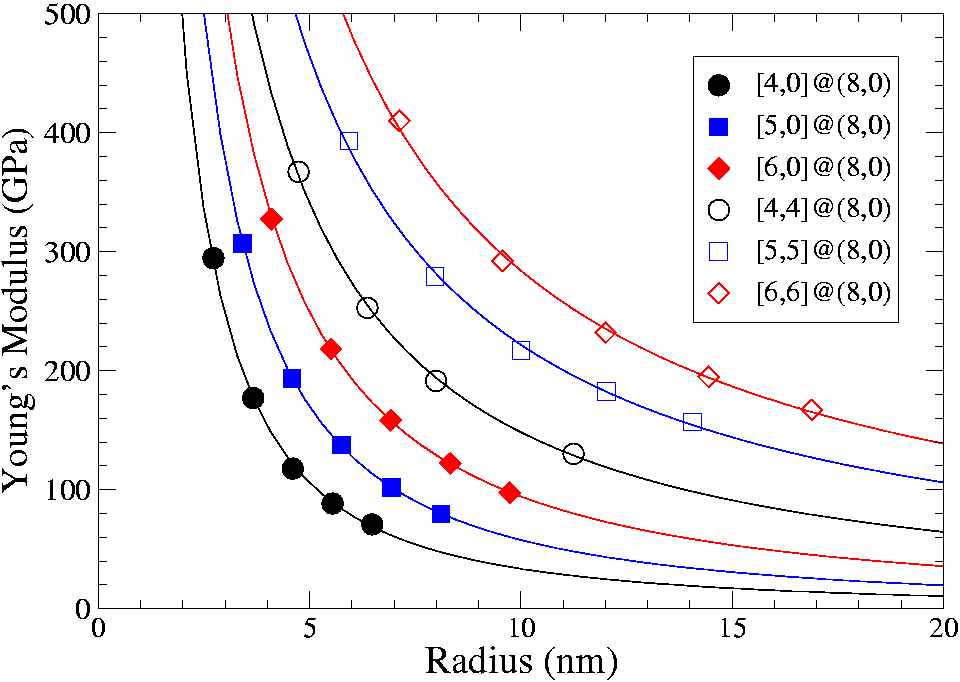}
\caption{Behavior of the dynamic Young's modulus ($E$) as a function of the super carbon nanotube radius ($R$). The points are results from atomistic calculations at 300 K and the lines represent the fitting of the points with the function $E \propto R^{-\beta}$. The $\beta$ values are presented in Table 1.}
\end{center}
\end{figure}

\begin{table}[ht]
\caption {Fitted values of the parameters $E_0$ and $\beta$ of the relation $E=E_0(R/R_0)^{-\beta}$. The value of $R_0$ was adopted as being the smallest radius for each set of [$N$,$M$]@(8,0) STs considered. The error bars correspond to the standard deviations which are smaller than 3\%.}
\vspace{0.05cm} \centering
%\footnotesize
\begin{tabular}{|c c c c|}
\hline
Structure  &  $E_0$ (GPa)  & $\beta$ & $R_0$ (nm)\\
\hline
$[4,0]@(8,0)$	& 290 $\pm$ 6 & 1.66 $\pm$ 0.04 & 2.72 \\
$[5,0]@(8,0)$	& 308 $\pm$ 2 & 1.56 $\pm$ 0.01 & 3.41 \\
$[6,0]@(8,0)$	& 329 $\pm$ 1 & 1.40 $\pm$ 0.01 & 4.10 \\
$[4,4]@(8,0)$	& 363 $\pm$ 5 & 1.20 $\pm$ 0.02 & 4.74 \\
$[5,5]@(8,0)$	& 387 $\pm$ 6 & 1.06 $\pm$ 0.03 & 5.93 \\
$[6,6]@(8,0)$	& 403 $\pm$ 6 & 1.03 $\pm$ 0.02 & 7.12 \\
\hline
\end{tabular}
\end{table}

Figure 3 shows the behavior of the tensile stress during tensile test for different STs. For comparison the results for the (8,0) SWCNT are also presented. The (8,0) SWCNT investigated here is predicted to present tensile strength of about 90 GPa and breaking strain of 28\% for 300 K in agreement with other theoretical simulations \cite{fracture1,susan}. Differences between the behaviors of the SWCNT and STs can be seen. Furthermore, there are also differences between zigzag and armchair STs. Zigzag STs present behaviors similar to the ones observed in brittle materials while armchair STs to ductile materials. It is worth to point out that STs are more flexible than the (8,0) SWCNT (i.e., (8,0) SWCNT is stiffer than STs) but in some cases (e.g., for zigzag STs) they show comparable strength. This aspect indicates that STs would behave as flexible, high-strength materials. 

The rupture in zigzag STs occurs at breaking strains close to the (8,0) SWCNT, i.e., 28$-$30\%. Snapshots of the deformation of a [4,0]@(8,0) ST are shown in Figure 4. The ST deformation is mainly due to angle changing rather than SWCNT stretching, as confirmed by the observed hyper-elasticity due to a fishing net like behavior (Figure 4). Similar behavior was also observed in simulations carried out for super-graphene sheets \cite{vitorprb}. During the tensile deformation (Figure 4 (a)-(d)) the angles between SWCNTs in the ST structures are changed and the stress concentration is mainly observed at the junctions (Figure 4 (b) and (c)). Romo-Herrera \textit{et al.} have also observed a stress concentration at junctions in their two- and three-dimensional proposed carbon nanotube networks \cite{terr}. The ST rupture occurs near the junctions close to the ST extremities where different SWCNTs break approximately at the same time, causing an abrupt decrease of the strain energy. Thus, due to the large changes in the angles between SWCNTs, the rupture happens before the occurrence of an effective stretching of SWCNTs. In fact, increasing the zigzag ST radius the breaking strain is shifted to a higher value than the one found in a zigzag ST with a smaller radius (Figure 3). At later stages of ST rupture, unraveling monoatomic chains are formed connecting SWCNT fragments, similar to ones observed in SWCNT fracture \cite{yakobson}.

The key points related to the different behaviors presented by zigzag and armchair STs are the junctions and their arrangement in the STs. We have associated the deformation behavior of STs with two main mechanisms. The first one is associated with the SWCNT stretching, dictated by the SWCNT mechanical property. The second one is due to the deformation at the junctions, i.e., changes in the angle of the junctions. These changes are relatively easier to occur than the SWCNT stretching due to the high value of the SWCNT Young's modulus. Consequently, for cases where the arrangement of junctions in the ST facilitates the deformation of the first type in a tensile test, the ST will present a tensile strength and breaking strain similar to the constituent SWCNT. It will not be exactly the same due to the effect of the second type of deformation. This is the case for the [4,0]@(8,0) STs. On the other hand, when the second type of deformation is privileged during the tensile test due to the arrangement of the junctions, the angle changes become more relevant than the SWCNT stretching, as in the case of the [4,4]@(8,0) STs.

The arrangement of SWCNTs to form a ST leads to the presence of large pores on the ST sidewalls. The size of these pores would be comparable to the typical size of proteins ($\sim$5$-$20 nm) \cite{coluci}. This property might make possible the use of STs as cavities and reservoirs. Furthermore, due to predicted high flexibility exhibited by STs (Figure 4 (a)-(d)), the shape and aperture of the pores presented by STs can be controlled by external mechanical loads, allowing a controllable access to the inner parts of zigzag STs through tensile strains. 

\begin{figure}
\begin{center}
\includegraphics[angle=0,width=120 mm]{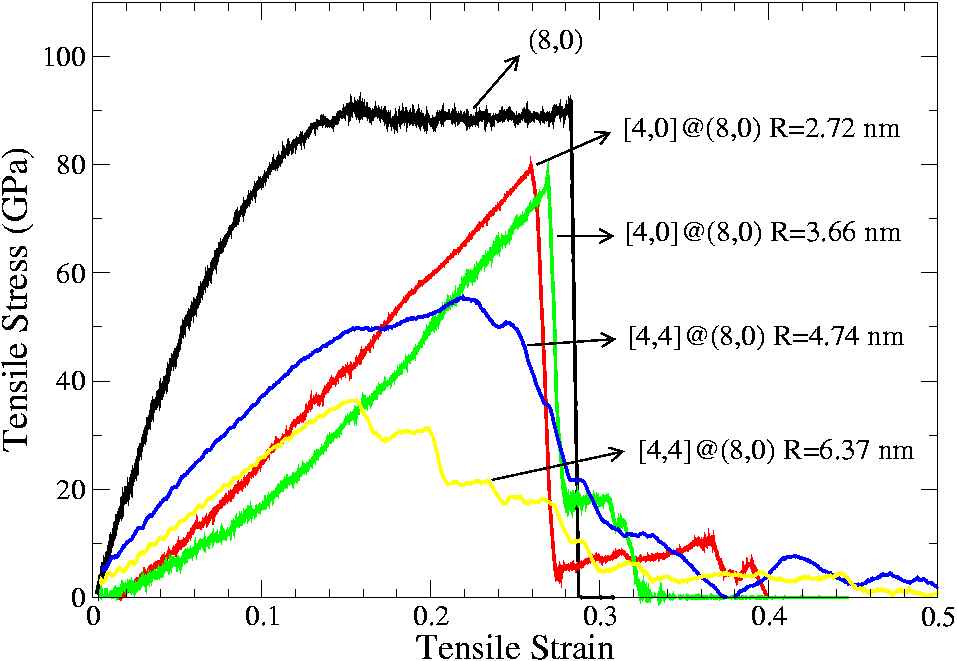}
\caption{Tensile failure behavior for different STs and for the (8,0) SWCNT at 300 K.}
\end{center}
\end{figure}

\begin{figure}
\begin{center}
\includegraphics[angle=0,width=120 mm]{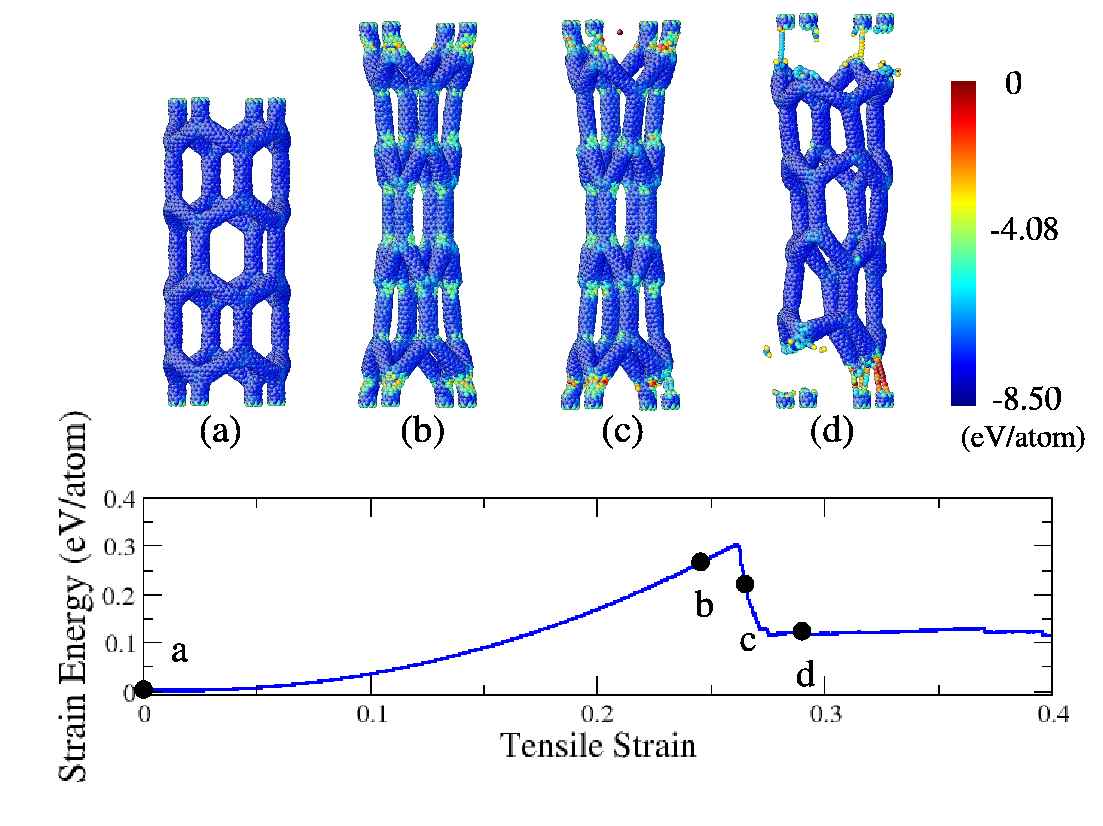}
\caption{Behavior of the strain energy as function of the tensile strain for the [4,0]@(8,0) ST with $R=$ 2.72 nm at 300 K. Snapshots of the simulation are presented for specific tensile strains. Atoms are colored according to their potential energy.} 
\end{center}
\end{figure}

For armchair STs we can also see a stress concentration mainly occurring at the junctions (Figure 5). On the other hand, the rupture of the armchair ST occurs not abruptly, as the zigzag case, but in small steps through a multi-step breaking of the junctions. This behavior can be seen in the strain energy evolution shown in Figure 5 between the points (c) and (d), where many decreasing steps of the strain energy are observed and associated with local ruptures on the ST structure. Contrarily to the zigzag case, the increasing of the armchair ST radius leads to a decreasing of the breaking strain value (Figure 3). As we have seen in Figure 2, more flexible STs are expected for larger radius STs. Such flexibility associated with the armchair ST geometry cause higher angle changes in tensile deformations that would lead to smaller breaking strain values.

\begin{figure}
\begin{center}
\includegraphics[angle=0,width=120 mm]{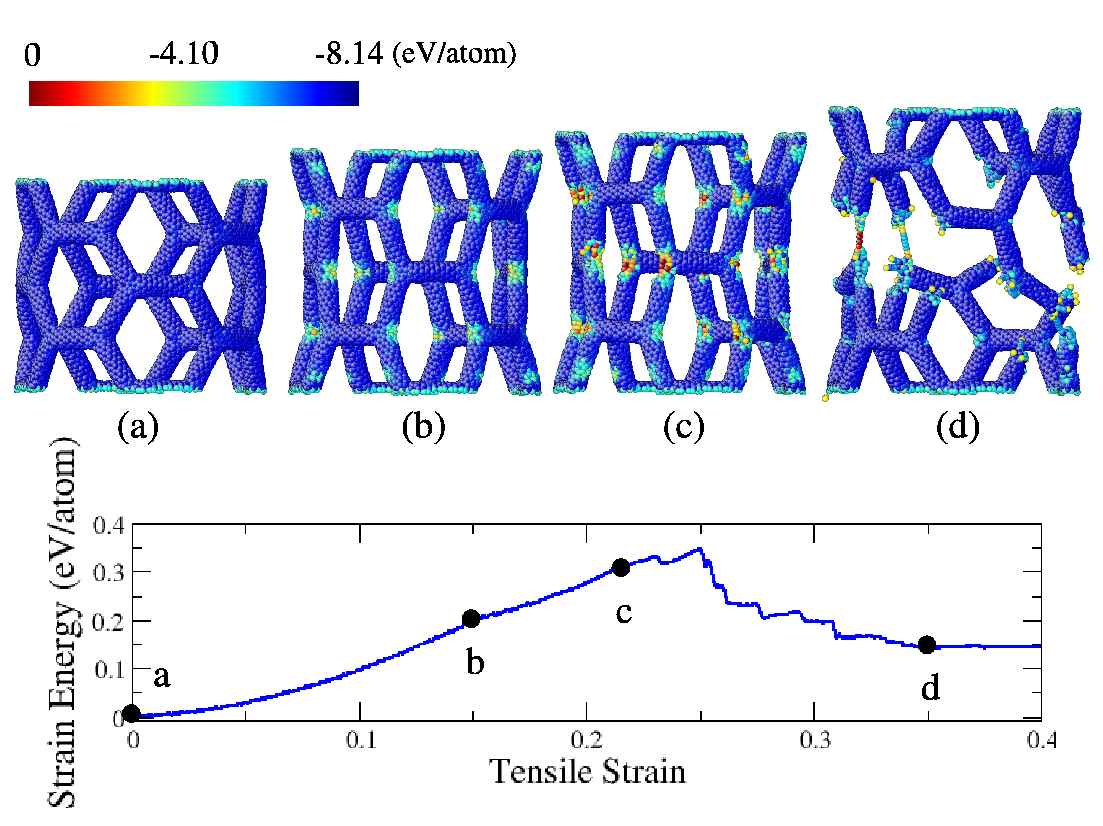}
\caption{Behavior of the strain energy as function of the tensile strain for the [4,4]@(8,0) ST with $R=$ 4.74 nm at 300 K. Snapshots of the simulation are presented for specific tensile strains. Atoms are colored according to their potential energy.} 
\end{center}
\end{figure}

For chiral STs, where a mixed ST structure between armchair and zigzag STs is presented, we have observed a mixed behavior. Figure 6 shows the evolution of strain energy of a chiral [4,1]@(8,0) ST. This behavior also shows the multi-step breaking as seen in armchair STs but the strain energy reduction in each step is larger than the one observed in armchair STs, resembling the case of zigzag STs. In this work we have limited ourselves to investigate the rupture of STs with typical radii values of 5 nm. For structures with larger values more flexible structures are expected. Further studies are necessary for establishing the behavior of STs with very large diameters and the role of the junctions in these cases. Studies along these lines are in progress.

\begin{figure}
\begin{center}
\includegraphics[angle=0,width=120 mm]{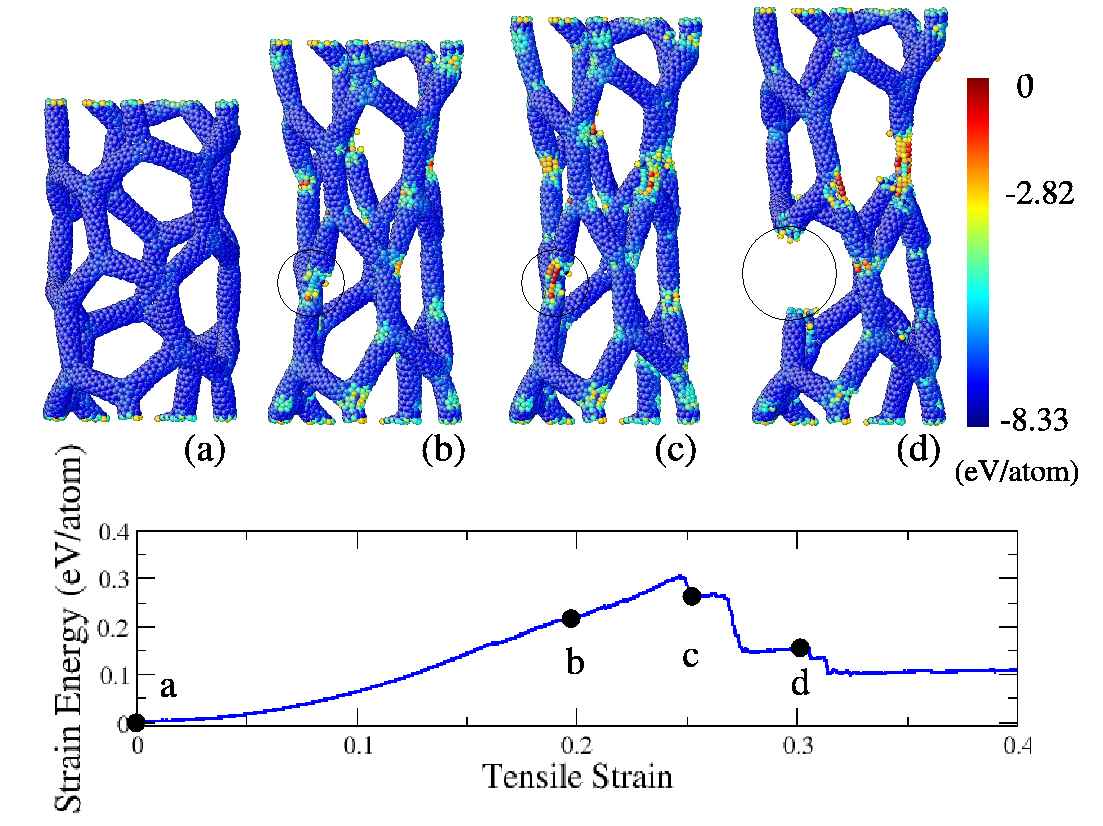}
\caption{Behavior of the strain energy as function of the tensile strain for the [4,1]@(8,0) ST with $R=$ 3.13 nm at 300 K. Snapshots of the simulation are presented for specific tensile strains. Atoms are colored according to their potential energy. Circles indicate the region which firstly breaks.} 
\end{center}
\end{figure}

Table 2 summarizes the results obtained for tensile strength and breaking strain for STs at 300 K and 1000 K. The STs present smaller tensile strength values than their constituent (8,0) SWCNTs. Similarly to the SWCNT, zigzag and armchair STs tensile strength and breaking strain also present smaller values for 1000 K. It is interesting to point out that, for the temperatures we have analyzed, the breaking strain values for zigzag STs are significantly temperature dependent. A reduction of the breaking strain value of $\sim$12\% is observed for zigzag STs in contrast to $\sim$43\% for the (8,0) SWCNT. Furthermore, the arrangement of SWCNTs in the zigzag STs seems to preserve the tensile strength presented by the constituent SWCNTs when the size of the structure (measured by its radius) is increased. This can be seen from the small reduction $\sim$15\% of the tensile strength when the structure radius increases by a factor $\sim$10. Interestingly, the SWCNT arrangement in chiral STs leads to an opposite behavior of tensile strength and breaking strain. Both quantities present larger values for 1000 K than 300 K.

\begin{table}[ht]
\caption {Atomistic results for the tensile strength $\sigma_m$ and breaking strain $\epsilon_m$ for 300 K and 1000 K. The values correspond to a single simulation. Approximated values for chiral STs are due to the approximations in the cross-sectional area determination.}
 \vspace{0.05cm} \centering
\begin{tabular}{|c c c c c c|}
 \hline
Structure  &  Radius (nm)  & \multicolumn{2}{c}{$\sigma_m$ (GPa)} & \multicolumn{2}{c|}{$\epsilon_m$} \\
& & 300 K & 1000 K & 300 K & 1000 K \\
\hline
$(8,0)$		& 0.31 & 92.6 & 88.7 & 0.28 & 0.16 \\
$[4,0]@(8,0)$  	& 2.72  & 79.2 & 63.1 & 0.26 & 0.23 \\
$[4,0]@(8,0)$  	& 3.66  & 77.9 & 68.3 & 0.27 & 0.25 \\
$[4,4]@(8,0)$	& 4.74  & 55.5 & 46.7 & 0.22 & 0.16 \\
$[4,4]@(8,0)$  	& 6.36  & 36.5 & 32.2 & 0.15 & 0.14   \\
$[4,1]@(8,0)$	& 3.13  & $\simeq$44 & $\simeq$54 & 0.24 & 0.27 \\
$[4,1]@(8,0)$  	& 4.20  & $\simeq$35 & $\simeq$49 & 0.24 & 0.28   \\
\hline
\end{tabular}
\end{table}

In order to quantify the energy absorption capacity of STs during impact loads we have calculated the fracture toughness ($K_c$), the fracture energy ($G_f$), and the dissipated energy per unit mass or toughness ($D_e$). The fracture energy (per unit area) is the area under the stress-strain curve multiplied by the initial ST length. The fracture toughness is defined as the square root of the product between the fracture energy and the Young's modulus. The dissipated energy was determined by calculating the area under the stress-strain curve divided by the material density (calculated using the volume comprised by the Connolly surface \cite{surface} obtained with a spherical probe of radius of 3 {\AA}). This quantity is more appropriate to characterize ductile materials, for which the dissipation is distributed on their volume; for brittle materials the fracture energy is more appropriate since the dissipation is localized on the cracked surface. Table 3 presents the values obtained for some STs.

\begin{table}[ht]
\caption {Atomistic results of the fracture toughness ($K_c$), fracture energy ($G_f$), and dissipated energy per unit mass ($D_e$) for 300 K and 1000 K.}
\vspace{0.05cm}
\centering
\begin{tabular}{|c c c c c c c c|}
 \hline
Structure  &  Radius (nm)  & \multicolumn{2}{c}{$K_c$ (MPa m$^{1/2}$)} & \multicolumn{2}{c}{$G_f$ (N/m)} &  \multicolumn{2}{c|}{$D_e$ (KJ/g)} \\
& & 300 K & 1000 K & 300 K & 1000 K & 300 K & 1000 K \\
\hline
$(8,0)$		& 0.31 & 25.6 & 17.1 & 876 & 395  & 10.7 & 4.8 \\
$[4,0]@(8,0)$  	& 2.72  & 6.8 & 5.9  & 160 & 119  & 5.5  & 4.1 \\
$[4,0]@(8,0)$  	& 3.66  & 5.8 & 5.1  & 188 & 143  & 4.9  & 3.8  \\
$[4,4]@(8,0)$	& 4.74  & 6.2 & 5.8  & 106 & 122  & 6.5  & 5.6  \\
$[4,4]@(8,0)$  	& 6.36  & 4.5 & 4.7  & 81 & 86  & 3.6  & 3.8  \\
$[4,1]@(8,0)$	& 3.13  & $\simeq$4 & $\simeq$6 & $\simeq$94 & $\simeq$177  & $\simeq$4  & $\simeq$8  \\
$[4,1]@(8,0)$  	& 4.20  & $\simeq$4 & $\simeq$4  & $\simeq$91 & $\simeq$113  & $\simeq$3  & $\simeq$4  \\\hline
\end{tabular}
\end{table}

The investigated STs presented fracture toughness of $\sim$5 MPa~m$^{1/2}$, fracture energy of $\sim$100 N/m, and dissipated energy during tensile strain of $\sim$5 KJ/g. These values are comparatively about 5, 8, and 2 times smaller than the ones presented by the (8,0) SWCNT at 300 K, respectively. We can also see that for STs these quantities are less sensitive to temperature increase than in the SWCNT case. These results suggest that STs can be used in applications that require the maintenance of mechanical properties when high temperature variations are expected, such as fuselage protection for spacecrafts. 

Recently carbon nanotube fibres comprising SWCNTs in a polymer matrix were produced showing a toughness of 570 J/g \cite{ray-nature} that is higher than the one presented by spider dragline silk (165 J/g \cite{spider}), Kevlar (33 J/g \cite{spider}) and graphite fibres (12 J/g \cite{graphite}). Using the estimated reduction ($\sim$50 \%) of toughness presented by STs compared to (8,0) SWCNTs at 300 K (Table 3) we can roughly estimate a real toughness value for STs as $\sim$570/2 = 285 J/g. According to Pugno \cite{pugno}, one of the most promising mechanical application of STs would be as multiscale fibre-reinforcements in tough matrices, mimicking bioinspired hierarchical materials, as bone, dentine (both having 7 hierarchical levels) or nacre (2 hierarchical levels). Note that a ST-based composite could be even more tough than the 
nanotube-based counterpart. This is mainly due to its multiscale topology, which could be able to activate toughening mechanisms and stop cracks at all the size-scales. This is observed in super-tough hierarchical biomaterials, such as bone, dentine or nacre. His analysis suggests that two-hierarchical levels STs would be the optimum for producing ``super''-composites (as exhibited by nacre in Nature). The optimization will be activated by the synergy between strong fibres and tough matrix, activating toughening (e.g., fibre pull-out) mechanisms. Therefore, our calculations, resulting in strengths of several GPa and toughness of $\sim$280 J/g, suggest that super-composites based on STs could be competitive with super-tough carbon nanotube fibres (strength of 1.8 GPa and toughness of 570 J/g \cite{ray-nature}), spider silk (strength of 1.8 GPa and toughness of 163 J/g \cite{ray-nature}), and Kevlar (strength of 3.6 GPa and toughness of 33 J/g \cite{spider}) even if longer fibres are expected to be weaker \cite{pugno4}. 

Atomistic simulations have been carried out only for (8,0) SWCNTs (ST$^{(0)}$), zigzag, armchair, and chiral STs (ST$^{(1)}$). However, the derived mechanical properties, such as Young's modulus and tensile strength, can be estimated for higher-order self-similar STs, following the procedure proposed in \cite{pugno}, i.e. introducing appropriate scaling laws. In particular, the Young's modulus $E$ for a ST of radius $R$ is expected to scale according to $E\approx E_0 (R/R_0)^{-\beta}$ \cite{pugno,pugno3}, as confirmed by our fittings reported in Figure 2 (see Table 1). Similarly, for the tensile strength $\sigma$ we expect $\sigma\approx \sigma_0 (R/R_0)^{D-2}$ where $D$ is the so-called fractal exponent. From our atomistic simulations we found for $D$ the values of 1.94 and 2.27 for [4,0]@(8,0) ST at 300 K and 1000 K or 0.57 and 0.74 for [4,4]@(8,0) ST at 300 K and 1000 K, respectively. For a self-similar ST of hierarchy $K$, the strength or Young's modulus $P$ scales, as a first approximation, as $P_K \approx P_0\varphi^K$ (the subscript 0 refers to a SWCNT), where $\varphi$ is the cross-sectional area fraction of ST of order $k-1$ in a ST of order $k$ (=1,$\ldots$,$K$) \cite{pugno}. For example, according to our atomistic simulations at 300 K, we find for the self-similar [4,0]@(8,0) ST$^{(2)}$, having first-order radius of 2.72 nm, a value for the strength of 68 GPa and for the Young's modulus of 57 GPa; for a self-similar [4,4]@(8,0) ST$^{(2)}$, having first-order radius of 4.74 nm, values for the strength of 33 GPa and for the Young's modulus of 89 GPa are computed. The scaling laws we are invoking here are of geometrical nature, thus they are valid, in principle, for all the hierarchical levels, if the related geometries are considered. Our derived predictions for the ST$^{(2)}$ are based on the hypothesis of geometrical self-similarity, i.e., the 
number and cross-sectional area fraction of SWCNTs inside the ST$^{(1)}$ are the 
same of those of ST$^{(1)}$s inside the studied ST$^{(2)}$.

\section{Summary and Conclusions}
The mechanical properties of the so-called `super' carbon nanotubes (STs) are investigated using classical molecular dynamics simulations based on reactive empirical bond-order potential. From tensile tests of impact loads, we have found that STs are more flexible than the SWCNT used to form them but, in some cases, they show comparable tensile strengths. The STs Young's modulus have been predicted to have an inverse proportionality with the ST radius. During tensile deformations the shape and aperture of pores in ST sidewalls can be modified providing a way to vary the accessible channels to the inner parts of STs through the application of mechanical loads. The ST rupture occurs basically at regions near the SWCNT junctions and it is influenced by the ST chirality. The investigated STs presented values of fracture toughness, fracture energy, and dissipated energy that are about 5, 8, and 2 times smaller than the ones presented by the constituent (8,0) SWCNT, respectively. Based on the predicted geometrical and mechanical properties, STs may represent new candidates for novel porous, flexible, and high-strength materials.

%\ack 
We acknowledge the financial support from THEO-NANO, the Instituto de
Nanosci\^encias and IMMP/CNPq, the Brazilian agencies FAPESP, CAPES, and CNPq.


\begin{thebibliography}{99}
\bibitem{surface}The Connolly surface is at the boundary between a probe and the atoms as the probe rolls over the surface that intersects with the van der Waals radii of the atoms in the structure.
\end{thebibliography}
\end{document}